\documentclass{webofc}
\usepackage[varg]{txfonts}   
\usepackage{adjustbox}

\usepackage{physics}\usepackage{xcolor}
\begin{document}
\title{Pion polarizabilities from a dispersive analysis of $\gamma\gamma \to \pi\pi$}
%
%

\author{\firstname{Viktoriia} \lastname{Ermolina}\inst{1}\fnsep\thanks{\email{vermolin@uni-mainz.de}}, 
\firstname{Igor} \lastname{Danilkin}\inst{1} \and
        \firstname{Marc} \lastname{Vanderhaeghen}\inst{1}
}

\institute{Institut f\"ur Kernphysik, Johannes Gutenberg Universit\"at, D-55099 Mainz, Germany}

\abstract{%
We present results for the charged and neutral pion polarizabilities, obtained through a dispersive analysis of photon-fusion reactions with two pions in the final state. This analysis is motivated by current and future measurements at COMPASS, JLab (Hall D), and the BESIII measurement of single-virtual photon-fusion reactions in the space-like region, offering insight into generalized pion polarizabilities. To improve predictions within an unsubtracted dispersion formalism with only the pion left-hand cut, one needs to consider the influence of heavier left-hand cuts. We parametrize the latter using an expansion in a suitably constructed conformal variable, capturing the analytical structure of the left-hand cuts. The coefficients in this expansion are determined from the matching with the Lagrangian-based approach, the Adler zero constraint for $\gamma\gamma \to \pi^0\pi^0$, and a fit to the cross-section data. The extension to the single-virtual case is also discussed.
}
\maketitle
\section{Introduction}\label{sec:1}

The $\gamma\gamma \to \pi\pi$ reaction is a promising way to study the electromagnetic characteristics of the pion, such as its polarizabilities. Utilizing dispersive techniques, one can analytically continue the $\gamma\gamma \to \pi\pi$ amplitude into the unphysical region ($t=m_\pi^2$ and $s=0$), and be close enough to the two-pion threshold where cross section data is available. More direct experimental extraction of pion polarizabilities is challenging and is typically carried out through various Primakoff processes \cite{Antipov:1982kz,COMPASS:2014eqi} or by radiative pion photoproduction on the proton \cite{Ahrens:2004mg}. Among these extractions, only the latest result for the charged pion polarizability from COMPASS \cite{COMPASS:2014eqi} confirms the chiral prediction \cite{Gasser:2006qa}. Additional measurements with higher statistics are planned or under discussion at Jefferson Lab's Hall D and at AMBER@CERN.

Our approach is based on the partial wave (p.w.) dispersion relations, which account for the analyticity and unitarity properties of the S-matrix. The unsubtracted dispersion relations with only pion-pole left-hand cut given in \cite{Colangelo:2017fiz,Colangelo:2017qdm} offer a reasonable description of the photon fusion cross-section data and predict the charged pion dipole polarizability in line with recent COMPASS measurements and ChPT. Meanwhile, the neutral pion dipole polarizability appears far from the two-loop ChPT result \cite{Gasser:2005ud}, exhibiting a different order of magnitude and a different sign. It is reasonable to assume that the neutral pion polarizability pattern will also hold for the single-virtual extension of the unsubtracted dispersion relation with only pion left-hand cut. Its impact on the $f_0(500)$ contribution to hadronic light-by-light (HLbL) scattering in $(g-2)_\mu$ remains to be shown.

The neutral pion dipole polarizability discrepancy can be potentially resolved by taking into account corrections from heavier left-hand cuts, which are dominated by the $\omega$ exchange \cite{Colangelo:2017fiz,Colangelo:2017qdm}. However, in the conventional scheme, this requires the introduction of subtraction constants that compromise the high-energy behavior of the p.w. amplitudes. The latter complicates utilizing cross-section data across an extended energy range. Instead, in order to model-independently capture the effect of the heavier left-hand cuts, we employ the conformal mapping technique. With experimental data in both neutral and charged channels, the unknown coefficients in the expansion can be directly fitted to the data. In the absence of precision data close to the threshold (which is the present case), we need to rely on a few assumptions - specifically, the Adler zero of $\gamma\gamma \to \pi^0\pi^0$ amplitude and a resonance exchange model for heavier left-hand cuts. 


\section{Formalism}\label{sec:2}
The kinematically unconstrained p.w. amplitudes rely on a decomposition of the hadronic tensor of $\gamma^{(*)}\gamma^{(*)}\to \pi\pi$ into a suitable set of invariant functions. For the real photon case, they are given by \cite{Garcia-Martin:2010kyn}
\begin{align}
\bar{h}^{(0)}_1=\frac{\bar{h}^{(0)}_{++}}{s}\,,\quad \bar{h}_1^{(2)}=\frac{\bar{h}^{(2)}_{+-}}{s(s-4m_\pi^2)}\,,\quad  \bar{h}_2^{(2)}=\frac{\bar{h}^{(2)}_{++}}{s^2(s-4m_\pi^2)}\,,
\end{align}
where $h_{\lambda_1\lambda_2}^{(J)}$ are the isospin p.w. helicity amplitudes (the isospin index $I$ is not explicitly written for simplicity), and "bar" stands for the Born subtracted amplitude, $\bar{h}\equiv h-h^{\text{Born}}\,$. For the single and double-virtual cases, the transformation to the kinematically unconstrained amplitudes is more complicated and detailed in \cite{Danilkin:2018qfn,Danilkin:2019opj} (also see \cite{Hoferichter:2019nlq}). The new set of amplitudes $h_i(s)$ contains only dynamical singularities, specifically the right- and left-hand cuts, and satisfies the standard dispersion relations. As pointed out in \cite{Garcia-Martin:2010kyn,Moussallam:2021dpk}, the solution to this dispersion relation can be expressed using different types of Muskhelishvili-Omn\`es (MO) representations, involving either only right-hand cut integrations or both, which are exactly equivalent \cite{Hoferichter:2019nlq}. However, since typically the left-hand cuts are not well known, the so-called modified MO representation for $\bar{h}^{(J)}_{i}(\Omega^{(J)})^{-1}$ \cite{Garcia-Martin:2010kyn} is a preferable choice and is given by
\begin{align}
\label{Rescattring:general_v1}
\bar{h}^{(J)}_{i}=\Omega^{(J)} \left(
-\int_{R}\frac{ds'}{\pi}\,\frac{\text{Im}\,(\Omega^{(J)}(s'))^{-1}\,{h}^{(J),\text{Born}}_{i}(s')}{s'-s}+\int_{L}
\frac{ds'}{\pi}\,\frac{(\Omega^{(J)}(s'))^{-1}\,\text{Im}\,{h}^{(J),\text{res}}_{i}(s')}{s'-s}
\right)\,,
\end{align}
where under the left-hand cut integral, we approximate the imaginary part of the Born-subtracted amplitude by the resonance exchanges \cite{Garcia-Martin:2010kyn}. In Eq.(\ref{Rescattring:general_v1}), $\Omega^{(J)}$ represents the Omn\`es functions, which in the single-channel approach are given as,
\begin{equation}\label{OmenesPhaseShift}
\Omega_I^{(J)}=\exp\left(\frac{s}{\pi}\int_{4m_\pi^2}^{\infty} \frac{d s'}{s'}\frac{\delta_{I}^{(J)}(s')}{s'-s}\right).
\end{equation}
Here the corresponding phase shifts for D-wave are taken from the Roy analysis \cite{Garcia-Martin:2011iqs}. For the S-wave amplitudes with isospin $I=0,2$, we utilize results obtained from a single-channel partial-wave dispersion relation \cite{Danilkin:2020pak} (see also mIAM \cite{GomezNicola:2007qj} and DIA \cite{Danilkin:2022cnj})

From the analysis of the radiative couplings, it becomes apparent that the dominant contribution arises from the $\omega$-exchange. Therefore, in \cite{Danilkin:2018qfn} only light vector mesons were considered. Contributions from other heavier resonances were effectively incorporated by slightly adjusting the  $VP\gamma$ coupling (which is defined in a Lagrangian governing the exchange of a vector meson), as detailed in \cite{Danilkin:2018qfn}:
\begin{align}\label{Eq:res=V}
h^{(J),\text{res}}_{i} \approx h^{(J),V}_{i},\quad   C_{\rho^{\pm,0}\pi^{\pm,0}\gamma}\overset{\text{SU(3)}}{\simeq} \frac{C_{\omega\pi^{0}\gamma}}{3} \approx  g^{\text{eff}}_{VP\gamma}\,.
\end{align}
The coupling $g^{\text{eff}}_{VP\gamma}$ is the only parameter used to dispersively describe the cross-section data around the $f_2(1270)$ resonance, resulting in $g^{\text{eff}}_{VP\gamma} =0.326(4)$ GeV$^{-1}$ \cite{Danilkin:2018qfn}. This value falls within $15\%$ of the PDG average $g^{\text{PDG}}_{VP\gamma}=0.38(2)$ GeV$^{-1}$ \cite{Workman:2022ynf}, thus justifying the approximation of left-hand cuts by vector mesons. Given that the dominant resonance contribution comes from $\omega$ exchange, forthcoming single-virtual BESIII data around the $f_2(1270)$ resonance \cite{Danilkin:2019mhd}, will enable us to better constrain the $\omega$ transition from factor.
This is particularly crucial for virtualities above 1 GeV$^2$, where the correct perturbative QCD asymptotic behavior may begin to play a significant role - a factor not presently accounted for in dispersive analyses \cite{Schneider:2012ez,Danilkin:2014cra}.

While it is possible to utilize the unsubtracted dispersion relation in the form of (\ref{Rescattring:general_v1}) with (\ref{Eq:res=V}) for the D-wave, this does not apply to the S-wave. The dispersive integral along the vector left-hand cuts is formally convergent, owing to the asymptotically bounded behavior of the S-wave Omn\`es functions and the imaginary part of the amplitude $h^{(0),V}_{++}$:
\begin{align}
\label{Rescattring:general_v3}
h^{(0)}_{++}&=h^{(0),\text{Born}}_{++}+\Omega^{(0)}\,\left(-\frac{s}{\pi}
\int_{R}\frac{ds'}{s'}\,\frac{\text{Im}\,(\Omega^{(0)}(s'))^{-1}\,{h}^{(0),\text{Born}}_{++}(s')}{s'-s}+\mathcal{LHC}\right)\\
\mathcal{LHC} &\equiv \frac{s}{\pi}\int_{L}
\frac{ds'}{s'}\,\frac{(\Omega^{(0)}(s'))^{-1}\,\text{Im}\,{h}^{(0),V}_{++}(s')}{s'-s}\,.\label{Eq:LHC}
\end{align}
However, the left-hand contribution acquires significant corrections from the integration over large negative $s$. Therefore, in \cite{Colangelo:2017fiz,Danilkin:2018qfn,Danilkin:2019opj,Hoferichter:2019nlq}, the results for the S-wave were produced based only on the pion-pole (Born) rescattering. To improve convergence under the dispersive integral over the left-hand cut, the conventional scheme necessitates the introduction of subtraction constants, compromising the high-energy behavior of the partial wave amplitudes. Instead, we propose utilizing the conformal mapping technique and approximating the contribution from heavier-than-pion left-hand cuts as follows
\begin{align}
\label{Rescattring:general_v4}
h^{(0)}_{++}=h^{(0),\text{Born}}_{++}+\Omega^{(0)}\,\Bigg(-\frac{s}{\pi}
\int_{R}\frac{ds'}{s'}\,\frac{\text{Im}\,(\Omega^{(0)}(s'))^{-1}\,{h}^{(0),\text{Born}}_{++}(s')}{s'-s}+\sum_{n=1}^{\infty} C_n\,\xi^n
\Bigg)\,,
\end{align}
with the property that $\xi(s=0)=0$, in order to ensure the soft-photon limit. The conformal mapping variable is chosen such that it maps the left-hand cut plane onto the unit circle. The form of $\xi$ depends on the cut structure of the reaction. For the real photon fusion reaction, all left-hand cuts lie on the real axis. Therefore, $\xi$ is given by
\begin{align}
\label{Eq:xi}
\xi=\frac{\sqrt{s-s_L}-\sqrt{s_E-s_L}}{\sqrt{s-s_L}+\sqrt{s_E-s_L}}\,,
\end{align}
with $s_E=0$. In Eq.(\ref{Eq:xi}) $s_L$ is the position of the closest left-hand cut branching point beyond the one pion exchange. Strictly speaking, it is given by the two-pion exchange in the $t$ and $u$ channels, $s^{(2\pi)}_L=-9m_\pi^2/4$. However, the two-pion discontinuity along these cuts concentrates around $t(u)=M_\rho^2$. A similar story applies to the three-pion left-hand cut $s^{(3\pi)}_L=-64\,m_\pi^2/9$ and $t(u)=M_\omega^2$, which provides an even better approximation due to the narrow width of the $\omega$-meson. Therefore, for our central result, we take
\begin{align}
\label{Eq:sL}
s_L \simeq s^{(V)}_L=-\frac{\left(m_V^2-m_\pi^2\right)^2}{m_V^2}\,.
\end{align}
For the single-virtual case with space-like photon virtuality, the left-hand cut consists of the two pieces $(-\infty,s_L^{(-)}]$ and $[s_{L}^{(+)},0]$. 
The cut $[s_{L}^{(+)},0]$ is tiny and very close to the physical region. Therefore, we propose explicitly incorporating its contribution evaluating the integral in Eq.(\ref{Eq:LHC}), while expanding the contribution from the dominant cut using the conformal variable with $s_L\simeq s_L^{(-)}$ and $s_E=-Q^2$. The latter is needed for the soft-photon limit.

Given the form of $\xi(s)$, the series in Eq.(\ref{Rescattring:general_v4}) truncated at any finite order is bounded asymptotically, consistent with the once-subtracted dispersion relation in (\ref{Rescattring:general_v3}). In general, the unknown coefficients $C_n$ can be determined from fitting cross-sectional data in the $\gamma\gamma^{(*)} \to \pi^+ \pi^-$ and $\gamma\gamma^{(*)} \to \pi^0 \pi^0$ channels. In the absence of very precise data in both channels, additional constraints from $\chi$PT and Lagrangian-based models for the heavier left-hand cuts can be used. Particularly, chiral dynamics requires the S-wave $\gamma\gamma \to \pi^0 \pi^0$ amplitude to have an Adler zero
\begin{align}
\label{Eq:constraint_1}
h^{n,(0)}_{++}(s_A)=0\,,\quad s_A=(1\pm 0.2)\, m_\pi^2\,,
\end{align}
where $s_A=m_\pi^2$ corresponds to NLO.
At NNLO, this position shifts to $s_A=1.175\,m_\pi^2$ \cite{Bellucci:1994eb}. Therefore, we consider a range similar to that in \cite{Dai:2014zta}. The Adler zero lies between the soft-photon zero and the threshold, thus accommodating the negative dipole polarizability for the neutral pion \cite{Dai:2016ytz}, which corresponds to a negative slope at $s=0$ for the corresponding amplitude. Note also that at NLO, the Adler zero for $\gamma^*\gamma^* \to \pi^0 \pi^0$ does not depend on the virtualities of the photons. Another constraint arises from the exchange of vector mesons. We propose the following matching condition
\begin{equation}\label{Eq:constraint_2}
    \left(\sum^\infty_{n=1} C_n \,\xi^n\right)''_{|s=0}=\mathcal{LHC}''_{|s=0}= \frac{2}{\pi}\int_{-\infty}^{s_L}
\frac{ds'}{s'^3}\,(\Omega^{(0)}(s'))^{-1}\,\text{Im}\,{h}^{(0),V}_{++}(s')\,,
\end{equation}
where the right-hand side of the above equation is essentially a sum rule (multiplied by a factor of 2) for the over subtracted $\mathcal{LHC}(s)$ at $s=0$. The integrand in Eq.(\ref{Eq:constraint_2}) is now suppressed by an additional power of $s$ in the denominator compared to Eq.(\ref{Eq:LHC}). For the single-virtual case Eq.(\ref{Eq:constraint_2}) modifies to
\begin{equation}\label{Eq:constraint_3}
    \left(\sum^\infty_{n=1} C_n(Q^2) \,\xi^n\right)''_{|s=-Q^2}=\mathcal{LHC}''_{|s=-Q^2}= \frac{2}{\pi}\int_{-\infty}^{s_L^{(-)}}
\frac{ds'}{(s'+Q^2)^3}\,(\Omega^{(0)}(s'))^{-1}\,\text{Im}\,{h}^{(0),V}_{++}(s')\,,
\end{equation}
with the corresponding MO representation
\begin{align}
\label{Rescattring:Q2}
h^{(0)}_{++}&=h^{(0),\text{Born}}_{++}+\Omega^{(0)}\,\Bigg(-\frac{s+Q^2}{\pi}
\int_{R}\frac{ds'}{s'+Q^2}\,\frac{\text{Im}\,(\Omega^{(0)}(s'))^{-1}\,{h}^{(0),\text{Born}}_{++}(s')}{s'-s}\nonumber\\
&+\frac{s+Q^2}{\pi}\int_{s_L^{(+)}}^{0}
\frac{ds'}{s'+Q^2}\,\frac{(\Omega^{(0)}(s'))^{-1}\,\text{Im}\,{h}^{(0),V}_{++}(s')}{s'-s}+\sum_{n=1}^{\infty} C_n(Q^2)\,\xi^n\Bigg)\,.
\end{align}

\begin{figure}[ht]
\centering
\includegraphics[width=0.60\textwidth]{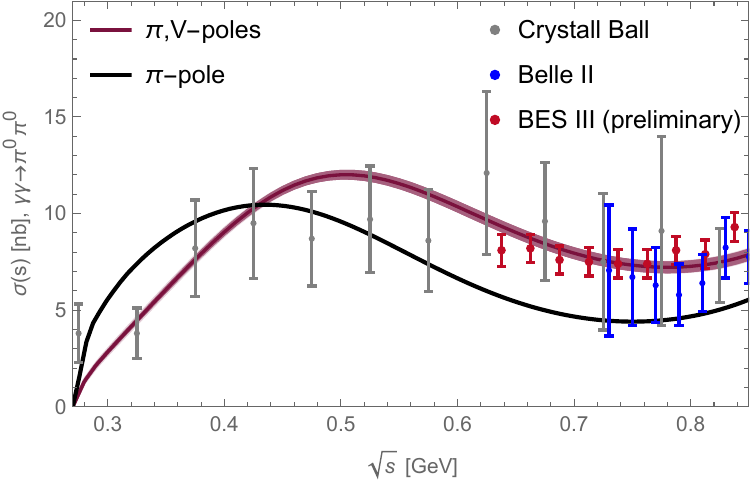} 
\caption{Integrated $\gamma\gamma\to\pi^0\pi^0$ cross-section with $\left|\cos{\theta}\right|\leq0.8$ corresponding to different left-hand cut contributions in the S-wave. The data are taken from \cite{CrystalBall:1990oiv,Belle:2009ylx,Kussner:2022dft,Kussner:2024ryb}.}
\label{fig:2}
\end{figure}

\section{Results}\label{sec:3}
Both the $\gamma\gamma\to\pi^0\pi^0$ and $\gamma\gamma\to\pi^+\pi^-$ channels have been experimentally measured. However, in the low-energy region, there is only old data from Crystal Ball \cite{CrystalBall:1990oiv} and Mark II \cite{Boyer:1990vu} with relatively large errorbars. For the $\pi^0\pi^0$ channel, starting from 0.64 GeV, there is also relatively precise data from Belle \cite{Belle:2009ylx} and preliminary BESIII \cite{Kussner:2022dft,Kussner:2024ryb}. In the current analysis of the cross-section data, we perform a single-channel description of the S-wave isoscalar channel, therefore limiting ourselves to energies below $\sqrt{s}=0.85$ GeV. The pion-pole post-diction of the cross-section, shown in Fig.\,\ref{fig:2}, corresponds to $\chi^2/N_{data}=6.5(1.5)$, where in brackets we show the result excluding preliminary BESIII data. The pion dipole and quadrupole polarizabilities are given in Table\,\ref{tab:1}, indicating a significant discrepancy, particularly for the dipole (and to a lesser extent, quadrupole) polarizability for the neutral pion. 

\begin{table}[h]
\caption{Results for dipole and quadrupole polarizabilities compared to Garcia-Martin et al. \cite{Garcia-Martin:2010kyn}, ChPT \cite{Gasser:2005ud,Gasser:2006qa} and data \cite{COMPASS:2014eqi}. The first error corresponds to the variation of $s_A$, while the second error indicates the uncertainties in the experimental cross section data. The central result corresponds to $g^{\text{eff}}_{VP\gamma}$. The third error represents the shift when using $g^{\text{PDG}}_{VP\gamma}$ instead of $g^{\text{eff}}_{VP\gamma}$, i.e. incorporating a potential slight variation in the effective coupling between the S and D waves.}
\begin{adjustbox}{width=1\textwidth}
\begin{tabular}{cccccc}
\hline\hline
& \multicolumn{3}{c}{Dispersive}  &  ChPT & Experiment \\
& \multicolumn{2}{c}{Present work} & Garcia-Martin et al. &  &  \\
& $\pi$ pole & $\pi, V$ poles & $\pi, V, A, T$ pole & NNLO & COMPASS\\ \hline
$\left(\alpha_1 - \beta_1\right)_{\pi^\pm}\left[10^{-4} \mbox{fm}^3\right] $ & $5.1$ & $2.4\,(4)(3)^{+1.0}_{-0.0}$ &  $4.7$ & $5.7\left(1.0\right)$ & $4.0\left(1.2\right)_{\mbox{stat}}\left(1.4\right)_{\mbox{sys}}$\\
$\left(\alpha_1 - \beta_1\right)_{\pi^0}\left[10^{-4} \mbox{fm}^3\right] $ & $8.4$ & $-1.3\,(3)(0)^{+0.0}_{-0.3}$ &  $-1.25\left(17\right)$ & $-1.9\left(2\right)$ & $-$\\
$\left(\alpha_2 - \beta_2\right)_{\pi^\pm}\left[10^{-4} \mbox{fm}^5\right] $ & $18.1$ & $16.5\,(4)(2)^{+2.1}_{-0.0}$ &  $14.7\left(2.1\right)$ & $16.2/21.6$ & $-$\\
$\left(\alpha_2 - \beta_2\right)_{\pi^0}\left[10^{-4} \mbox{fm}^5\right] $ & $24.8$ & $30.0\,(4)(3)^{+4.3}_{-0.0}$ &  $32.1\left(2.1\right)$ & $37.6\left(3.3\right)$ & $-$\\
\hline\hline
\end{tabular}
\end{adjustbox}
\label{tab:1}  
\end{table}

In the scheme with heavier than pion-pole left-hand cuts, we truncate the series in (\ref{Rescattring:general_v4}) at $n_{max}=3$, i.e., take the first two dominant terms into account in both isospin channels. This corresponds to 4 unknown parameters. If one fixes these parameters using ChPT values given in Table~\ref{tab:1}, then the description of the neutral cross-section improves to $\chi^2/N_{data}=1.0(0.7)$, showing the consistency between the ChPT values and experimental data in a dispersive approach. A similar result was shown in \cite{Hoferichter:2011wk} using the Roy-Steiner equations. We next want to explore the possibility of extracting polarizabilities directly from the current real photon data. Clearly, we found out that the current cross-section data cannot constrain 4 parameters from a direct fit, therefore we rely on a few assumptions - specifically, the Adler zero of $\gamma\gamma \to \pi^0\pi^0$ amplitude (see Eq.(\ref{Eq:constraint_1})) and resonance exchange model for heavier left-hand cuts (see Eq.(\ref{Eq:constraint_2})). One free parameter left in the expansion scheme can then be adjusted to the available cross-section data. The results of the fit correspond to $\chi^2/(N_{data}-1)=0.8(0.6)$ and given in Table\,\ref{tab:1} and Fig.\,\ref{fig:2}. The obtained charged dipole polarizability turned out to be somewhat smaller than the ChPT result, but still consistent with the COMPASS measurement. Clearly, $\gamma\gamma\to\pi^+\pi^-$ data, planned at Jefferson Lab's Hall D, are needed to obtain the charged polarizability more accurately.



The forthcoming step is to extend this analysis to the coupled-channel cases $(\gamma\gamma\to\pi\pi/K\Bar{K},\gamma\gamma\to\pi\eta/K\Bar{K})$ and access kaon polarizabilities. Furthermore, the current analysis can be applied to the finite virtuality of one of the photons $Q^2$ in combination with future data from BESIII.

\section*{Acknowledgments}
This work was supported by the Deutsche Forschungsgemeinschaft (DFG, German Research
Foundation) within the Research Unit [Photon-photon interactions in the Standard Model and beyond, Projektnummer 458854507 - FOR 5327].

\bibliography{sn-bibliography}

\begin{thebibliography}{29}

\bibitem{Antipov:1982kz}
Y.M. Antipov et~al., Phys. Lett. B \textbf{121}, 445 (1983)

\bibitem{COMPASS:2014eqi}
C.~Adolph et~al. (COMPASS), Phys. Rev. Lett. \textbf{114}, 062002 (2015)

\bibitem{Ahrens:2004mg}
J.~Ahrens et~al., Eur. Phys. J. A \textbf{23}, 113 (2005)

\bibitem{Gasser:2006qa}
J.~Gasser, M.A. Ivanov, M.E. Sainio, Nucl. Phys. B \textbf{745}, 84 (2006)

\bibitem{Colangelo:2017fiz}
G.~Colangelo, M.~Hoferichter, M.~Procura, P.~Stoffer, JHEP \textbf{04}, 161 (2017)

\bibitem{Colangelo:2017qdm}
G.~Colangelo, M.~Hoferichter, M.~Procura, P.~Stoffer, Phys. Rev. Lett. \textbf{118}, 232001 (2017)

\bibitem{Gasser:2005ud}
J.~Gasser, M.A. Ivanov, M.E. Sainio, Nucl. Phys. B \textbf{728}, 31 (2005)

\bibitem{Garcia-Martin:2010kyn}
R.~Garcia-Martin, B.~Moussallam, Eur. Phys. J. C \textbf{70}, 155 (2010)

\bibitem{Danilkin:2018qfn}
I.~Danilkin, M.~Vanderhaeghen, Phys. Lett. B \textbf{789}, 366 (2019)

\bibitem{Danilkin:2019opj}
I.~Danilkin, O.~Deineka, M.~Vanderhaeghen, Phys. Rev. D \textbf{101}, 054008 (2020)

\bibitem{Hoferichter:2019nlq}
M.~Hoferichter, P.~Stoffer, JHEP \textbf{07}, 073 (2019)

\bibitem{Moussallam:2021dpk}
B.~Moussallam, Eur. Phys. J. C \textbf{81}, 993 (2021)

\bibitem{Garcia-Martin:2011iqs}
R.~Garcia-Martin, R.~Kaminski, J.R. Pelaez, J.~Ruiz~de Elvira, F.J. Yndurain, Phys. Rev. D \textbf{83}, 074004 (2011)

\bibitem{Danilkin:2020pak}
I.~Danilkin, O.~Deineka, M.~Vanderhaeghen, Phys. Rev. D \textbf{103}, 114023 (2021)

\bibitem{GomezNicola:2007qj}
A.~Gomez~Nicola, J.R. Pelaez, G.~Rios, Phys. Rev. D \textbf{77}, 056006 (2008)

\bibitem{Danilkin:2022cnj}
I.~Danilkin, V.~Biloshytskyi, X.L. Ren, M.~Vanderhaeghen, Phys. Rev. D \textbf{107}, 074021 (2023)

\bibitem{Workman:2022ynf}
R.L. Workman, Others (Particle Data Group), PTEP \textbf{2022}, 083C01 (2022)

\bibitem{Danilkin:2019mhd}
I.~Danilkin, C.F. Redmer, M.~Vanderhaeghen, Prog. Part. Nucl. Phys. \textbf{107}, 20 (2019)

\bibitem{Schneider:2012ez}
S.P. Schneider, B.~Kubis, F.~Niecknig, Phys. Rev. D \textbf{86}, 054013 (2012)

\bibitem{Danilkin:2014cra}
I.V. Danilkin et~al., Phys. Rev. D \textbf{91}, 094029 (2015)

\bibitem{Bellucci:1994eb}
S.~Bellucci, J.~Gasser, M.E. Sainio, Nucl. Phys. B \textbf{423}, 80 (1994)

\bibitem{Dai:2014zta}
L.Y. Dai, M.R. Pennington, Phys. Rev. D \textbf{90}, 036004 (2014)

\bibitem{Dai:2016ytz}
L.Y. Dai, M.R. Pennington, Phys. Rev. D \textbf{94}, 116021 (2016)

\bibitem{CrystalBall:1990oiv}
H.~Marsiske et~al. (Crystal Ball), Phys. Rev. D \textbf{41}, 3324 (1990)

\bibitem{Belle:2009ylx}
S.~Uehara et~al. (Belle), Phys. Rev. D \textbf{79}, 052009 (2009)

\bibitem{Kussner:2022dft}
M.~K\"u\ss{}ner, Ph.D. thesis, Ruhr U., Bochum (2022)

\bibitem{Kussner:2024ryb}
M.~K\"u\ss{}ner (BESIII), plenary talk at MESON2023, EPJ Web Conf. \textbf{291}, 01002 (2024)

\bibitem{Boyer:1990vu}
J.~Boyer et~al., Phys. Rev. D \textbf{42}, 1350 (1990)

\bibitem{Hoferichter:2011wk}
M.~Hoferichter, D.R. Phillips, C.~Schat, Eur. Phys. J. C \textbf{71}, 1743 (2011)

\end{thebibliography}

\end{document}